\documentclass[prd,twocolumn,showpacs,superscriptaddress,nofootinbib,floatfix,showkeys,10pt]{revtex4-2}

\usepackage{graphicx}
\usepackage{amsmath}
\usepackage{bm}
\usepackage{yhmath}
\usepackage{mathtools}
\usepackage{wasysym}
\usepackage{color,xcolor}
\usepackage{subfigure}
\usepackage{cases}
\usepackage{times}
\usepackage{dcolumn,booktabs,bm}
\usepackage{slashed}
\usepackage{amsfonts,amssymb,stmaryrd,latexsym}
\usepackage{mathrsfs}
\usepackage{textcomp}
\usepackage{multirow}
\usepackage{cancel}
\usepackage{array}
\usepackage{enumerate,enumitem}
\usepackage{orcidlink}
\usepackage{dashrule}
\usepackage{multirow}
\usepackage{mdframed}
\usepackage{hyperref}

\newcommand{\etal}{\textit{et al}.}

\definecolor{mycolor}{RGB}{45,48,146}
\hypersetup{colorlinks=true,citecolor=mycolor,urlcolor=mycolor,linkcolor=mycolor}

\allowdisplaybreaks[4]
\newcounter{cnt}
\makeatletter
\let\oldhypertarget\hypertarget
\renewcommand{\hypertarget}[2]{%
  \oldhypertarget{#1}{#2}%
    \protected@write\@mainaux{}{%
        \string\expandafter\string\gdef
          \string\csname\string\detokenize{#1}\string\endcsname{#2}%
    }%
  }
\newcommand{\myhyperlink}[1]{%
  \hyperlink{#1}{\csname #1\endcsname}%
  }
\makeatother

\newcommand{\clabel}[2][]{#2}
\newcommand{\change}[1]{#1}
\begin{document}

%%%%%%%%%%%%%%%%open the reply mode%%%%%%%%%%%%%%%%%%
%\onecolumngrid
%\input{reply_a.tex}
%\newpage
%\setcounter{page}{0}
%%%%%%%%%%%%%%%%open the reply mode%%%%%%%%%%%%%%%%%%
\title{Strong decays of the isovector-scalar $D^\ast\bar{D}^\ast$ hadronic molecule}

\author{Jin-Cheng Deng}
\affiliation{College of Physics Science and Technology, Hebei University, Baoding 071002, China}

\author{Bo Wang\,\orcidlink{0000-0003-0985-2958}}\email{wangbo@hbu.edu.cn}
\affiliation{College of Physics Science and Technology, Hebei University, Baoding 071002, China}
\affiliation{Hebei Key Laboratory of High-precision Computation and Application of Quantum Field Theory, Baoding, 071002, China}
\affiliation{Hebei Research Center of the Basic Discipline for Computational Physics, Baoding, 071002, China}

\begin{abstract}
We adopt the effective Lagrangian approach to study the strong decays of the $1^-(0^{++})$ $D^\ast\bar{D}^\ast$ molecular state [denoted as $T_{\psi0}^a(4010)$ according to the LHCb naming convention] through triangle diagrams. The decay channels include the open-charm $D\bar{D}$, and the hidden-charm $\eta_c\pi$, $J/\psi\rho$, and $\chi_{c1}\pi$. The coupling between the $T_{\psi0}^a(4010)$ and its constituents $D^\ast\bar{D}^\ast$ is obtained by solving for the residue of the scattering $T$-matrix at the pole. Our calculations yield a total width of few tens MeV for the $T_{\psi0}^a(4010)$ state using three different form factors, with its main decay channels being $\eta_c\pi$ and $\chi_{c1}\pi$. The $X(4100)$ and $X(4050)$ have similar masses and widths, with both masses being close to the $D^\ast\bar{D}^\ast$ threshold. Additionally, their decay final states are consistent with those of the $T_{\psi0}^a(4010)$. Therefore, it is likely that they represent the same state and both potentially correspond to the $T_{\psi0}^a(4010)$.
We suggest that future experiments focus on searching for the $T_{\psi0}^a(4010)$ signal in the final states $\eta_c\pi^-$, $\chi_{c1}\pi^-$ and $D^0D^-$ of the $B^0\to\eta_c\pi^-K^+$, $\chi_{c1}\pi^- K^+$ and $D^0D^-K^+$ processes, respectively, as well as further investigating its resonance parameters with Flatt\'e-like formula.
\end{abstract}

%\pacs{12.39.Fe, 12.39.Hg, 14.40.Nd, 14.40.Rt}
\maketitle

\section{Introduction}\label{sec:intro}
 In 2018, the LHCb Collaboration reported a resonance state, denoted as $X(4100)^-$~\cite{Workman:2022ynf}, in the invariant mass spectrum of $\eta_c\pi^-$ through the process $B^0\to\eta_c\pi^-K^+$ with a significance exceeding three standard deviations~\cite{LHCb:2018oeg}. The measured mass and width of the resonance are $m=4096\pm20^{+18}_{-22}$ MeV and $\Gamma=152\pm58^{+60}_{-35}$ MeV, respectively. However, the current data sample is insufficient to determine whether its spin-parity quantum numbers $J^P$ are $0^+$ or $1^-$. It can be inferred that this charged structure is an isovector state, with quark composition $c\bar{c}d\bar{u}$. Like the well-established states such as $Z_c(3900)$~\cite{BESIII:2013ris} and $Z_c(4020)$~\cite{BESIII:2013ouc}, the $X(4100)$ is also evidently an exotic charmoniumlike state.
 
 Several studies have explored the properties of the $X(4100)$ resonance. For instance, Wang \etal~employed~the QCD sum rule calculations, supporting the interpretation that $X(4100)$ is a scalar tetraquark~\cite{Wang:2018ntv}.~Wu \etal~also favored the assignment of $X(4100)$ as a $0^{++}$ tetraquark based on calculations using the chromomagnetic interaction model~\cite{Wu:2018xdi}. In contrast, Voloshin proposed that $X(4100)$ is a hadrocharmonium---a structure formed by a compact charmonium $\eta_c$ coupled to light quark excitations with pion quantum number through a QCD analogue of van der Waals force~\cite{Voloshin:2018vym}. In Ref.~\cite{Zhao:2018xrd}, Zhao proposed two possible interpretations for the $X(4100)$: (i) arising from the rescattering effect of the S-wave $D^\ast\bar{D}^\ast$, (ii) a genuine P-wave resonance of the $D^\ast\bar{D}^\ast$ system. Cao \etal~\cite{Cao:2018vmv}, on the other hand, proposed that the $X(4100)^-$ is the charge conjugate of the $X(4050)^+$ [with mass and width $m=4051\pm14^{+20}_{-41}$ MeV and $\Gamma=82^{+21+47}_{-17-22}$ MeV] previously observed by the Belle Collaboration in the invariant mass spectrum of $\chi_{c1}\pi^+$ via the decay $\bar{B}^0\to\chi_{c1}\pi^+K^-$~\cite{Belle:2008qeq}. Chen also indicated that the $X(4100)^-$ and $X(4050)^+$ could be the same state~\cite{Chen:2023iee}. For the other related works, see Refs.~\cite{Sundu:2018nxt,Mohammadi:2022zim}.
 
 The proximity of the $X(4100)$ resonance to the $D^\ast\bar{D}^\ast$ threshold naturally raises question about its potential connections with other resonances near the $D\bar{D}^\ast$ and $D^\ast\bar{D}^\ast$ thresholds. These resonances include the well-established $X(3872)$ and $Z_c(3900)$ near the $D\bar{D}^\ast$ threshold, and the $Z_c(4020)$ near the $D^\ast\bar{D}^\ast$ threshold. Despite ongoing debates regarding their exact nature, with interpretations ranging from hadronic molecules to compact tetraquarks and kinematic effects, the molecular picture has emerged as the most prevalent framework for understanding these states~\cite{Meng:2022ozq}. Within the molecular paradigm, their existence is tightly linked to the interactions of the $D^\ast\bar{D}^{(\ast)}$ systems with different quantum numbers.

 Since the discovery of $X(3872)$~\cite{Belle:2003nnu}, a series of studies have been conducted to investigate the interactions of the $D^\ast\bar{D}^{(\ast)}$ systems and explore the possible molecular states~\cite{Liu:2009qhy,Nieves:2012tt,Hidalgo-Duque:2012rqv,Sun:2012zzd,Li:2012cs,Guo:2013sya,Aceti:2014uea,Albaladejo:2015dsa,Baru:2016iwj,Liu:2019stu,Wang:2020dko,Wang:2020dgr,Xin:2022bzt,Peng:2023lfw}. Inspired by the recent observation of the $T_{cs0}(2900)$~\cite{LHCb:2020bls} and $T_{c\bar{s}0}^a(2900)$~\cite{LHCb:2022sfr}, we proposed to connect different hadronic molecules from the quark-level perspective in our recent works~\cite{Wang:2023hpp,Wang:2023eng,Wang:2024riu}. If $X(3872)$ and $Z_c(3900)$ are the bound and virtual states of the isoscalar and isovector $D\bar{D}^\ast$ system, respectively, we can obtain a virtual state in the $1(0^+)$ $D^\ast K^\ast$ system~\cite{Wang:2023hpp}, which is very likely to correspond to the newly observed $T_{c\bar{s}0}^a(2900)$ by the LHCb~\cite{LHCb:2022sfr}. At the same time, we found that the interaction of the $1^-(0^{++})$ $D^\ast \bar{D}^\ast$ system is equal to that of the $1(0^+)$ $D^\ast K^\ast$ system, and its interaction strength is about twice that of the $1^+(1^{+-})$ $D^\ast\bar{D}^{(\ast)}$ system (see Table IV of Ref.~\cite{Wang:2023hpp}, in which the interaction is dominated by the $\tilde{c}_a$ term). That is to say, if $Z_c(3900)/Z_c(4020)$ are molecular states of the $1^+(1^{+-})$ $D\bar{D}^{\ast}/D^\ast\bar{D}^{\ast}$ systems, then there must also exist a $1^-(0^{++})$ $D^\ast \bar{D}^\ast$ molecular state. Since its interaction is stronger, it corresponds to a bound state rather than a virtual state. The mass range of the $1^-(0^{++})$ $D^\ast \bar{D}^\ast$ state we obtain is $(4007.2$$-$$4016.7)$ MeV~\cite{Wang:2023hpp}. %The LHCb-observed $X(4100)$~\cite{LHCb:2018oeg} and Belle-observed $X(4050)$ are good candidates for the $1^-(0^{++})$ $D^\ast \bar{D}^\ast$ molecule predicted in Ref.~\cite{Wang:2023hpp}.
 
 The $D^\ast \bar{D}^\ast$ molecule with $1^-(0^{++})$ will decay into the open-charm $D\bar{D}$, and hidden-charm $\eta_c\pi$, $J/\psi\rho$ and $\chi_{c1}\pi$ channels. Therefore, further experimental measurements of the mass and width of $X(4100)/X(4050)$ with higher precision, as well as its partial decay widths to the $D\bar{D}$, $\eta_c\pi$, $J/\psi\rho$ and $\chi_{c1}\pi$ channels, will be very helpful to distinguish whether $X(4100)$~\cite{LHCb:2018oeg} and $X(4050)$~\cite{Belle:2008qeq} are the same state, and whether they are both the $1^-(0^{++})$ $D^\ast \bar{D}^\ast$ molecule. In this work, we employ the triangle diagram approach to calculate its strong decays, providing useful information such as total width, partial widths, and branching fractions/ratios. The coupling constant to its constituents $D^\ast \bar{D}^\ast$ will be determined from the residue of the scattering $T$-matrix at the pole~\cite{Meng:2021jnw}. The triangle diagram approach, also known as the hadron loop mechanism, has been widely applied to deal with the strong transitions of highly excited charmonium (bottomonium) states via the dipion ($\eta$, $\phi$, $\omega$) emissions~\cite{Guo:2010ak,Chen:2011qx,Chen:2011zv,Wang:2015xsa,Wang:2016qmz,Huang:2017kkg}, as well as the decays~\cite{Xiao:2019mvs,Lin:2019qiv} and productions~\cite{Guo:2013zbw,Wu:2023rrp} of hadronic molecules.
 
 The structure of this article is arranged as follows: In Sec.~\ref{sec:effl}, we will present the effective Lagrangians, coupling constants, and decay amplitudes. In Sec.~\ref{sec:nums}, we will provide our results and related discussions. In Sec.~\ref{sec:sum}, we will summarize this work.
 
\section{Effective Lagrangians, coupling constants, and decay amplitudes}\label{sec:effl}

\subsection{Effective Lagrangians}

\begin{figure}[!ht]
\begin{centering}
%\vspace{0.3cm}
    %\setlength{\abovecaptionskip}{0.01cm}
    %\setlength{\belowcaptionskip}{-0.4cm}
    \scalebox{1.0}{\includegraphics[width=\linewidth]{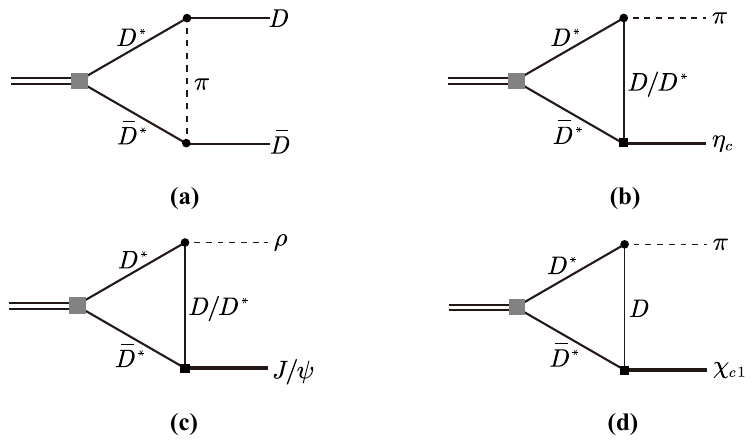}}
    \caption{The triangle loop diagrams for the strong decays of the $T_{\psi0}^a(4010)$ state. The double line, single line, and dashed line denote the $T_{\psi0}^a(4010)$, the $D/D^\ast$, and the light mesons $\pi/\rho$, respectively, while the charmonia $\eta_c$, $J/\psi$ and $\chi_{c1}$ are denoted by the thick lines. The contributions arising from the G-parity transformation are implied in figures (b), (c) and (d).\label{fig:decays}}
\end{centering}
\end{figure}

Following the recent LHCb naming convention~\cite{Gershon:2022xnn}, the $D^\ast \bar{D}^\ast$ molecule with $1^-(0^{++})$ quantum numbers should be named $T_{\psi0}^a(4010)$, with the mass in parentheses coming from our prediction~\cite{Wang:2023hpp}. Throughout the subsequent sections, we will refer to it simply as $T_{\psi0}^a$.~The decay processes of this state to $D\bar{D}$, $\eta_c\pi$, $J/\psi\rho$ and $\chi_{c1}\pi$ are shown in Fig.~\ref{fig:decays}, from which it can be seen that we need the following effective Lagrangians.

\begin{itemize}
\item [(i)] The Lagrangians for $T_{\psi0}^a D^\ast\bar{D}^\ast$ coupling.---The $T_{\psi0}^a$ couples to $D^\ast\bar{D}^\ast$ via the S-wave interaction, thus their Lagrangian can be written as
\begin{eqnarray}\label{eq:L0}
\mathcal{L}_{0}&=&g_{0}\tilde{P}^{\ast\dagger}_\mu T_{\psi0}^a P^{\ast\dagger\mu}+\mathrm{H.c.},
\end{eqnarray}
where $\tilde{P}^{\ast\dagger}_\mu=(\bar{D}^{\ast0\dagger},D^{\ast-\dagger})_\mu$, $P^{\ast\dagger}_\mu=(D^{\ast0\dagger},D^{\ast+\dagger})_\mu^T$. The $T_{\psi0}^a$ here denotes its isospin triplet,
\begin{eqnarray}
T_{\psi0}^a&=&\left[\begin{array}{cc}
\frac{1}{\sqrt{2}}T_{\psi0}^{a0} & T_{\psi0}^{a+}\\
T_{\psi0}^{a-} & -\frac{1}{\sqrt{2}}T_{\psi0}^{a0}
\end{array}\right].
\end{eqnarray}
The extraction of the coupling constant $g_0$ from the residual of $D^\ast \bar{D}^\ast$ scattering $T$-matrix will be demonstrated in Sec.~\ref{sec:coupg0}.
\item [(ii)] The Lagrangians for $D^\ast D^{(\ast)}\pi$ and $D^\ast D^{(\ast)}\rho$ couplings.---The corresponding Lagrangians within the superfield representations are given by~\cite{Meng:2022ozq,Casalbuoni:1996pg}
\begin{eqnarray}
  \mathcal{L}_\pi &=& g_b\langle\mathcal{H}\gamma^\mu\gamma^5u_\mu\bar{\mathcal{H}}\rangle + g_b\langle\bar{\tilde{\mathcal{H}}}\gamma^\mu\gamma^5u_\mu\tilde{\mathcal{H}}\rangle,\label{eq:Lpi} \\
  \mathcal{L}_\rho&=&i\beta\left\langle \mathcal{H}v^{\mu}(\Gamma_{\mu}-\rho_{\mu})\bar{\mathcal{H}}\right\rangle +i\lambda\left\langle \mathcal{H}\sigma^{\mu\nu}F_{\mu\nu}\bar{\mathcal{H}}\right\rangle \nonumber\\
  && + i\beta\left\langle \bar{\tilde{\mathcal{H}}}v^{\mu}(\Gamma_{\mu}-\rho_{\mu})\tilde{\mathcal{H}}\right\rangle +i\lambda\left\langle \bar{\tilde{\mathcal{H}}}\sigma^{\mu\nu}F_{\mu\nu}\tilde{\mathcal{H}}\right\rangle,\label{eq:Lrho}
\end{eqnarray}
where the notation $\langle x\rangle$ denotes the trace of $x$ in spinor space, and
\begin{eqnarray}
\mathcal{H}&=&\frac{1+\slashed{v}}{2}\left(P_{\mu}^{\ast}\gamma^{\mu}+iP\gamma^{5}\right),\\
\tilde{\mathcal{H}}&=&\left(\tilde{P}_{\mu}^{\ast}\gamma^{\mu}+i\tilde{P}\gamma^{5}\right)\frac{1-\slashed{v}}{2},
\end{eqnarray}
with $\bar{\mathcal{H}}=\gamma^0\mathcal{H}^\dagger\gamma^0$, $\bar{\tilde{\mathcal{H}}}=\gamma^0\tilde{\mathcal{H}}^\dagger\gamma^0$. The $P^{(\ast)}$ and $\tilde{P}^{(\ast)}$ are respectively given by $P^{(\ast)}=(D^{(\ast)0},D^{(\ast)+})$ and $\tilde{P}^{(\ast)}=(\bar{D}^{(\ast)0},D^{(\ast)-})^T$. $v^\mu$ denotes the four velocity of heavy mesons. The coupling constant $|g_b|=0.59$ is extracted from the partial decay width of the $D^\ast$ meson, e.g., the process $D^{\ast+}\to D^0\pi^+$~\cite{Workman:2022ynf}. Meanwhile, the axial-vector current $u_\mu$ and chiral connection $\Gamma_\mu$ are defined as
\begin{eqnarray}
u_\mu=\frac{i}{2}\left\{\xi^\dagger,\partial_\mu\xi\right\},\qquad \Gamma_\mu=\frac{1}{2}\left[\xi^\dagger,\partial_\mu\xi\right],
\end{eqnarray}
with 
\begin{eqnarray}
\xi^2=U=\exp\left(\frac{i\varphi}{f_\pi}\right), \varphi&=&\left[\begin{array}{cc}
\pi^0 & \sqrt{2}\pi^+\\
\sqrt{2}\pi^- & -\pi^0
\end{array}\right],
\end{eqnarray}
and the pion decay constant $f_\pi=92.4$ MeV. The antisymmetric tensors $\sigma^{\mu\nu}$ and $F_{\mu\nu}$ are given by
\begin{eqnarray}
\sigma^{\mu\nu}&=&\frac{i}{2}\left[\gamma^\mu,\gamma^\nu\right],\\
 F_{\mu\nu}&=&\partial_{\mu}\rho_{\nu}-\partial_{\nu}\rho_{\mu}+\left[\rho_{\mu},\rho_{\nu}\right],
\end{eqnarray}
with
\begin{eqnarray}
\rho_{\mu}=i\frac{g_{V}}{\sqrt{2}}V_\mu,\qquad V_\mu&=&\left[\begin{array}{cc}
\frac{\omega+\rho^0}{\sqrt{2}} & \rho^+\\
\rho^- & \frac{\omega-\rho^0}{\sqrt{2}}
\end{array}\right]_\mu,
\end{eqnarray}
and the constants $g_V=m_\rho/(\sqrt{2}f_\pi)=5.9$, $\beta=0.9$, and $\lambda=0.56$ GeV$^{-1}$~\cite{Casalbuoni:1996pg}.
\item [(iii)] The Lagrangians for $\eta_c D^{(\ast)}\bar{D}^\ast$, $J/\psi D^{(\ast)}\bar{D}^\ast$, and $\chi_{c1}D\bar{D}^\ast$ couplings.---For heavy quarkonium, such as charmonium, the heavy quark flavor symmetry is badly broken, while the heavy quark spin symmetry still holds. Therefore, charmonia with the same orbital angular momentum $L$ but different spins $S$ form multiplets. For example, the S-wave ($L=0$) spin doublet $\mathcal{J}$ containing the $J/\psi$ and $\eta_c$ are given by
\begin{eqnarray}
\mathcal{J}=\frac{1+\slashed{v}}{2}\left(\psi^\mu\gamma_\mu+i\eta_c\gamma_5\right)\frac{1-\slashed{v}}{2},
\end{eqnarray}
while the P-wave quartet $\mathcal{J}^{\prime\mu}$ including the $\chi_{c2}$, $\chi_{c1}$, $\chi_{c0}$ and $h_c$ are
\begin{eqnarray}
\mathcal{J}^{\prime\mu}&=&\frac{1+\slashed{v}}{2}\bigg(\chi_{c2}^{\mu\alpha}\gamma_\alpha+\frac{1}{\sqrt{2}}\epsilon^{\mu\alpha\beta\delta}v_\alpha\gamma_\beta\chi_{c1\delta}\nonumber\\
&&+\frac{1}{\sqrt{3}}(\gamma^\mu-v^\mu)\chi_{c0}+h_c^\mu\gamma_5\bigg)\frac{1-\slashed{v}}{2}.
\end{eqnarray}
Then the corresponding Lagrangians are given by~\cite{Colangelo:2003sa}
\begin{eqnarray}
\mathcal{L}_1&=&g_1 \langle \mathcal{J}^{\prime\mu}\bar{\tilde{\mathcal{H}}}\gamma_\mu\bar{\mathcal{H}}\rangle+\mathrm{H.c.},\label{eq:Lg1}\\
\mathcal{L}_2&=&g_2 \langle \mathcal{J}\bar{\tilde{\mathcal{H}}}\overleftrightarrow{\slashed{\partial}}\bar{\mathcal{H}}\rangle+\mathrm{H.c.},\label{eq:Lg2}
\end{eqnarray}
in which the coupling constants $g_1$ and $g_2$ are respectively related to the decay constants of $\chi_{c0}$ and $J/\psi$ with the vector meson dominance model~\cite{Colangelo:2003sa}.
\begin{eqnarray}
g_1=-\sqrt{\frac{m_{\chi_{c0}}}{3}}\frac{1}{f_{\chi_{c0}}},\qquad g_2=\frac{\sqrt{m_\psi}}{2m_D f_\psi},
\end{eqnarray}
with $f_{\chi_{c0}}=510\pm40$ MeV being estimated from the QCD sum rule~\cite{Colangelo:2002mj}, while the $f_\psi$ can be determined through the electromagnetic decay of $J/\psi$, i.e., with the decay width of $J/\psi\to e^+e^-$~\cite{Workman:2022ynf}
\begin{eqnarray}
\Gamma[J/\psi\to e^+e^-]=C_\psi\frac{4\pi}{3}\frac{\alpha^2}{m_\psi}f_\psi^2,
\end{eqnarray}
where $C_\psi=\frac{4}{9}$ is the charge square of charm quark, $\alpha=\frac{1}{137}$ is the fine-structure constant. One can obtain that $f_\psi=415$ MeV.
\end{itemize}

\subsection{Determination of the coupling constant $g_0$}\label{sec:coupg0}

In this section, we will demonstrate how to determine the coupling constant $g_0$ in Eq.~\eqref{eq:L0} using the residue of the scattering $T$-matrix at the pole. To begin, we consider the scattering of particles $A$ and $B$ (with masses $m_1$ and $m_2$, respectively) within the framework of field theory. Assuming that their interaction is strong enough, allowing for the formation of bound states, we must treat their scattering in a nonperturbative manner, which means using the Bethe-Salpeter (BS) equation,
\begin{eqnarray}\label{eq:bse}
T^{\mathrm{FT}}=V^{\mathrm{FT}}+V^{\mathrm{FT}}G^{\mathrm{FT}}T^{\mathrm{FT}},
\end{eqnarray}
where the superscript ``$\mathrm{FT}$" means that we are working in a relativistic field theoretical approach. The two-body relativistic propagator $G^{\mathrm{FT}}$ reads
\begin{eqnarray}\label{eq:relg}
G^{\mathrm{FT}}&=&i\int\frac{d^{4}q}{(2\pi)^{4}}\frac{1}{q^{2}-m_{1}^{2}+i\epsilon}\frac{1}{(P-q)^{2}-m_{2}^{2}+i\epsilon}\nonumber\\
&=&\int\frac{q^{2}dq}{(2\pi)^{2}}\frac{\omega_{1}+\omega_{2}}{\omega_{1}\omega_{2}}\frac{1}{P_{0}^{2}-(\omega_{1}+\omega_{2})^{2}+i\epsilon},
\end{eqnarray}
where $P=p_1+p_2$, with $p_1$ and $p_2$ the momentum of $A$ and $B$ particles, respectively, and $P_0$ denotes the center of mass energy of $AB$ system. The result in the second line of Eq.~\eqref{eq:relg} is obtained with the residue theorem. The $\omega_1$ and $\omega_2$ are given by
\begin{eqnarray}
\omega_{1}^{2}=\bm{q}^{2}+m_{1}^{2},\qquad \omega_{2}^{2}=\bm{q}^{2}+m_{2}^{2}.
\end{eqnarray}

Then, we perform a nonrelativistic reduction
\begin{eqnarray}
\omega_{i}&=&m_{i}+\frac{\bm{q}^{2}}{2m_{i}}+\dots,\quad (i=1,2),\\
\frac{\omega_{1}+\omega_{2}}{\omega_{1}\omega_{2}}&\approx&\frac{1}{m_1}+\frac{1}{m_2}=\frac{1}{\mu},
\end{eqnarray}
and 
\begin{eqnarray}
\frac{1}{P_{0}^{2}-(\omega_{1}+\omega_{2})^{2}}&=&\frac{1}{P_{0}-(\omega_{1}+\omega_{2})}\frac{1}{P_{0}+(\omega_{1}+\omega_{2})}\nonumber\\
&\simeq&\frac{1}{P_{0}-m_{1}-m_{2}-\frac{\bm{q}^{2}}{2\mu}}\frac{1}{2P_{0}}.
\end{eqnarray}
With the nonrelativistic reduction, the $G^{\mathrm{FT}}$ becomes
\begin{eqnarray}\label{eq:gftg}
G^{\mathrm{FT}}&=&\frac{\pi}{\mu\sqrt{s}}G,
\end{eqnarray}
where 
\begin{eqnarray}\label{eq:nonrelg}
G&=&\int\frac{q^{2}dq}{(2\pi)^{3}}\frac{1}{\sqrt{s}-m_{1}-m_{2}-\frac{\bm{q}^{2}}{2\mu}}
\end{eqnarray}
represents the nonrelativistic two-body propagator. Note that we have replaced the $P_0$ with $\sqrt{s}$, where $s=(p_1+p_2)^2$ is the invariant mass of the $AB$ system, and it equals to the $P_0^2$ in the center of mass frame.

For nonrelativistic scattering, such as the interaction of the $AB$ system near the threshold energy, we can use the Lippmann-Schwinger (LS) equation to handle it, i.e,
\begin{eqnarray}\label{eq:lse}
t=v+vGt,
\end{eqnarray}
where $G$ is given in Eq.~\eqref{eq:nonrelg}.

The integral equation in Eqs.~\eqref{eq:bse} and \eqref{eq:lse} can be transformed into algebraic equations with, e.g., the on-shell factorization approach~\cite{Oller:1997ti},
\begin{eqnarray}
T^{\mathrm{FT}}&=&(1-V^{\mathrm{FT}}G^{\mathrm{FT}})^{-1}V^{\mathrm{FT}},\\
t&=&(1-vG)^{-1}v.
\end{eqnarray}
It can be seen that the position of the pole in the $T$-matrix is determined by the product of $V^{\mathrm{FT}}$ ($v$) and $G^{\mathrm{FT}}$ ($G$). The following condition ensures that the pole position remains unchanged after the reduction,
\begin{eqnarray}
V^{\mathrm{FT}}G^{\mathrm{FT}}=vG.
\end{eqnarray}
With the relation in Eq.~\eqref{eq:gftg}, one can obtain that $V^{\mathrm{FT}}=\frac{\mu\sqrt{s}}{\pi}V$. Similarly, the $T^{\mathrm{FT}}$ and $t$ will satisfy the same relationship, i.e.,
\begin{eqnarray}
T^{\mathrm{FT}}=\frac{\mu\sqrt{s}}{\pi}t.
\end{eqnarray}

Now, we assume that there exists a pole with the squared mass $s_R$ in the $T$-matrix, e.g., corresponding to a bound state. Then, in the vicinity of this pole, the $T$-matrix can be approximately written as
\begin{eqnarray}\label{eq:tgft}
T^{\mathrm{FT}}=\frac{(g^{\mathrm{FT}})^2}{s-s_R},
\end{eqnarray}
where $g^{\mathrm{FT}}$ represents the coupling constant of the bound state $R$ to its component $AB$. The relation in Eq.~\eqref{eq:tgft} allows us to extract the coupling constant from the residue of $T$-matrix,
\begin{eqnarray}\label{eq:gft1}
(g^{\mathrm{FT}})^2&=&\lim_{s\to s_{R}}(s-s_{R})T^{\mathrm{FT}}\nonumber\\
&=&\lim_{s\to s_{R}}(\sqrt{s}+\sqrt{s_{R}})\frac{\mu\sqrt{s}}{\pi}(\sqrt{s}-\sqrt{s_{R}})t\nonumber\\
&=&\frac{2\mu s_{R}}{\pi}\lim_{s\to s_{R}}(\sqrt{s}-\sqrt{s_{R}})t,
\end{eqnarray}
and
\begin{eqnarray}\label{eq:rest}
\lim_{s\to s_{R}}(\sqrt{s}-\sqrt{s_{R}})t=\lim_{s\to s_{R}}\left[\frac{d}{d\sqrt{s}}t^{-1}(\sqrt{s})\right]^{-1},
\end{eqnarray}
where we have used the l'H\^opital rule.

We adopt sharp cutoff to regularize the integral in Eq.~\eqref{eq:nonrelg}. For the bound state case, i.e., $\sqrt{s}<m_{1}+m_{2}$, its expression reads
\begin{eqnarray}
G_\Lambda(\sqrt{s})=\frac{\mu}{4\pi^{3}}\left[\gamma_b\arctan\left(\frac{\Lambda}{\gamma_b}\right)-\Lambda\right],
\end{eqnarray}
where $\gamma_b=\sqrt{2\mu(m_{1}+m_{2}-\sqrt{s})}$ denotes the binding momentum, and $\Lambda$ is the cutoff parameter.

If the effective potential $v$ does not depend on energy, as in the case given in Ref.~\cite{Wang:2023hpp}. Then the Eq.~\eqref{eq:rest} will equal to 
\begin{eqnarray}\label{eq:gft2}
g^{\prime2}&=&\lim_{s\to s_{R}}\left[-\frac{d}{d\sqrt{s}}G_\Lambda(\sqrt{s})\right]^{-1}\nonumber\\
&=&\frac{\gamma_b}{\frac{\mu^{2}}{4\pi^{3}}\left[\arctan\left(\frac{\Lambda}{\gamma_b}\right)-\frac{\gamma_b\Lambda}{\gamma_b^{2}+\Lambda^{2}}\right]}.
\end{eqnarray}
The $g^{\mathrm{FT}}$ is obtained by combining the results in Eqs.~\eqref{eq:gft1}-\eqref{eq:gft2}. For the small binding case $\gamma_b\to 0$ and using $\Lambda\to\infty$, one can get $g^{\mathrm{FT}}=4\sqrt{\pi\gamma_b s_R/\mu}$. This is consistent with the relation used in Ref.~\cite{Meng:2021jnw}.

\begin{figure}[!ht]
\begin{centering}
%\vspace{0.3cm}
    %\setlength{\abovecaptionskip}{0.01cm}
    %\setlength{\belowcaptionskip}{-0.4cm}
    \scalebox{0.5}{\includegraphics[width=\linewidth]{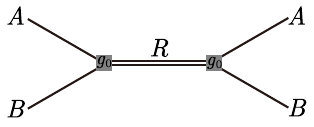}}
    \caption{The elastic scattering of $A$ and $B$ particles, where the $s$-channel is saturated with their bound state $R$.\label{fig:matchmt}}
\end{centering}
\end{figure}

To obtain the $g_0$ in Lagrangian~\eqref{eq:L0}, we still need to perform one final step:
We constructed the Lagrangian~\eqref{eq:L0} to describe the coupling between the bound state $T_{\psi0}^a$ and its components $D^\ast\bar{D}^\ast$. Using this Lagrangian, we can describe the elastic scattering process shown in Fig.~\ref{fig:matchmt} (replacing $A$, $B$, and $R$ with $D^\ast$, $\bar{D}^\ast$, and $T_{\psi0}^a$, respectively). The form of the scattering amplitude after nonrelativistic reduction is
\begin{eqnarray}\label{eq:elasm}
i\mathcal{M}=-i\frac{g_{0}^{2}}{s-s_{R}}(\bm{\varepsilon}_{1}\cdot\bm{\varepsilon}_{2})(\bm{\varepsilon}_{1}^{\dagger}\cdot\bm{\varepsilon}_{2}^{\dagger}),
\end{eqnarray}
where the $s_{R}$ here denotes the squared mass of $T_{\psi0}^a$, and $\bm{\varepsilon}_{1}$ ($\bm{\varepsilon}_{1}^{\dagger}$) and $\bm{\varepsilon}_{2}$ ($\bm{\varepsilon}_{2}^{\dagger}$) represent the polarization vectors of the initial (final) state $D^\ast$ and $\bar{D}^\ast$, respectively. Eq.~\eqref{eq:tgft} is also equivalent to an elastic scattering process shown in Fig.~\ref{fig:matchmt}. The scattering $T$-matrix is obtained in the partial wave basis. To match Eq.~\eqref{eq:elasm} with Eq.~\eqref{eq:tgft} equivalently, the scattering amplitude in Eq.~\eqref{eq:elasm} should also be projected onto the partial wave basis (S-wave). With the spin transition operators~\cite{Wang:2019ato}, the $(\bm{\varepsilon}_{1}\cdot\bm{\varepsilon}_{2})(\bm{\varepsilon}_{1}^{\dagger}\cdot\bm{\varepsilon}_{2}^{\dagger})$ equals to $(\bm{S}_1\cdot\bm{S}_2)^2-1$, with $\bm{S}_1$ ($\bm{S}_2$) the spin operator of the vector meson $D^\ast$ ($\bar{D}^\ast$). Then the Eq.~\eqref{eq:elasm} in the partial wave basis for the $^1S_0$ case is given by
\begin{eqnarray}
i\mathcal{M}_{^1S_0}=-i\frac{3g_{0}^{2}}{s-s_{R}}.
\end{eqnarray}
Finally, the matching condition 
\begin{eqnarray}
\mathcal{M}_{^1S_0}=-\frac{(g^{\mathrm{FT}})^2}{s-s_R}
\end{eqnarray}
yields
\begin{eqnarray}
g_0^2=\frac{2\mu_{D^\ast\bar{D}^\ast}m_{T_{\psi0}^a}^2}{3\pi}g^{\prime2},
\end{eqnarray}
where $\mu_{D^\ast\bar{D}^\ast}$ and $m_{T_{\psi0}^a}$ denote the reduced mass of $D^\ast\bar{D}^\ast$ and the mass of $T_{\psi0}^a$, respectively. The expression for $g^{\prime2}$ is given in Eq.~\eqref{eq:gft2}. The range of $g_0$ is
\begin{eqnarray}\label{eq:g0}
g_0\in[2.9,8.0]~\mathrm{GeV},
\end{eqnarray}
where we have used $\sqrt{s_R}=m_{T_{\psi0}^a}\in[4007.2,4016.7]$ MeV, and $\Lambda=0.4$ GeV~\cite{Wang:2023hpp}.

\subsection{Decay amplitudes and form factors}

By expanding the Lagrangians in Eqs.~\eqref{eq:L0}, \eqref{eq:Lpi}, \eqref{eq:Lrho}, \eqref{eq:Lg1} and \eqref{eq:Lg2}, we can obtain the vertices required for each Feynmann diagram shown in Fig.~\ref{fig:decays}. Considering the decay of $T_{\psi0}^a$ to the final states $f_2f_1$ through the process $$T_{\psi0}^a\to D^\ast(p_1+q)\bar{D}^\ast(p_2-q)[M(q)]\to f_2(p_2)f_1(p_1),$$ where the quantity in parenthesis represents the momentum of each particle, and the notation $[M]$ means that the $D^\ast\bar{D}^\ast$ transitions to $f_2f_1$ via exchanging the meson $M$. The amplitudes for each process from the diagrams in Fig.~\ref{fig:decays} are
\begin{widetext}
\begin{eqnarray}
i\mathcal{M}_{\textbf{(a)}}^{[\pi]}&=&C_{\textbf{(a)}}^{[\pi]}\int\frac{d^{4}q}{(2\pi)^{4}}\frac{g_{\mu}^{\alpha}-(p_{1}+q)_{\mu}(p_{1}+q)^{\alpha}/m_{D^{\ast}}^{2}}{(p_{1}+q)^{2}-m_{D^{\ast}}^{2}+i\epsilon} \frac{g^{\mu\beta}-(p_{2}-q)^{\mu}(p_{2}-q)^{\beta}/m_{D^{\ast}}^{2}}{(p_{2}-q)^{2}-m_{D^{\ast}}^{2}+i\epsilon}\frac{q_{\alpha}q_{\beta}}{q^{2}-m_{\pi}^{2}+i\epsilon},\label{eq:ddpi}\\
i\mathcal{M}_{\textbf{(b)}}^{[D]}&=&iC_{\textbf{(b)}}^{[D]}\int\frac{d^{4}q}{(2\pi)^{4}}\frac{g_{\alpha}^{\nu}-(p_{1}+q)_{\alpha}(p_{1}+q)^{\nu}/m_{D^{\ast}}^{2}}{(p_{1}+q)^{2}-m_{D^{\ast}}^{2}+i\epsilon} \frac{g^{\alpha\mu}-(p_{2}-q)^{\alpha}(p_{2}-q)^{\mu}/m_{D^{\ast}}^{2}}{(p_{2}-q)^{2}-m_{D^{\ast}}^{2}+i\epsilon}\frac{(p_{2}-2q)_{\mu}p_{1\nu}}{q^{2}-m_{D}^{2}+i\epsilon},\\
i\mathcal{M}_{\textbf{(b)}}^{[D^\ast]}&=&iC_{\textbf{(b)}}^{[D^{\ast}]}\int\frac{d^{4}q}{(2\pi)^{4}}\frac{g_{\delta\omega}-(p_{1}+q)_{\delta}(p_{1}+q)_{\omega}/m_{D^{\ast}}^{2}}{(p_{1}+q)^{2}-m_{D^{\ast}}^{2}+i\epsilon} \frac{g_{\alpha}^{\delta}-(p_{2}-q)^{\delta}(p_{2}-q)_{\alpha}/m_{D^{\ast}}^{2}}{(p_{2}-q)^{2}-m_{D^{\ast}}^{2}+i\epsilon}\frac{g_{\gamma\nu}-q_{\gamma}q_{\nu}/m_{D^{\ast}}^{2}}{q^{2}-m_{D^{\ast}}^{2}+i\epsilon} \nonumber\\ &&\times\left[\epsilon^{\mu\nu\alpha\beta}\epsilon^{\rho\omega\gamma\lambda}q_{\lambda}(p_{2}-2q)_{\mu}p_{1\rho}p_{2\beta}\right],\\
i\mathcal{M}_{\textbf{(c)}}^{[D]}&=&iC_{\textbf{(c)}}^{[D]}\int\frac{d^{4}q}{(2\pi)^{4}}\frac{g_{\gamma\delta}-(p_{1}+q)_{\gamma}(p_{1}+q)_{\delta}/m_{D^{\ast}}^{2}}{(p_{1}+q)^{2}-m_{D^{\ast}}^{2}+i\epsilon} \frac{g_{\alpha}^{\delta}-(p_{2}-q)^{\delta}(p_{2}-q)_{\alpha}/m_{D^{\ast}}^{2}}{(p_{2}-q)^{2}-m_{D^{\ast}}^{2}+i\epsilon}\frac{1}{q^{2}-m_{D}^{2}+i\epsilon}\nonumber\\
&&\times\left[\epsilon^{\mu\nu\alpha\beta}\epsilon^{\kappa\omega\gamma\lambda}q_{\lambda}(2q-p_{2})_{\mu}p_{1\kappa}p_{2\beta}\varepsilon_{\boldsymbol{\psi}\nu}^\dagger\varepsilon_{\boldsymbol{\rho}\omega}^\dagger\right],\\
i\mathcal{M}_{\textbf{(c)}}^{[D^\ast]}&=&iC_{\textbf{(c)}}^{[D^\ast]}\int\frac{d^{4}q}{(2\pi)^{4}}\frac{(p_{2}-2q)_{\mu}}{\left[(p_{1}+q)^{2}-m_{D^{\ast}}^{2}+i\epsilon\right]\left[(p_{2}-q)^{2}-m_{D^{\ast}}^{2}+i\epsilon\right]\left(q^{2}-m_{D^{\ast}}^{2} +i\epsilon\right)}\nonumber\\ &&\times\sum_{\lambda_{1},\lambda_{2},\lambda_{3}}\varepsilon_{1\delta}^{\dagger}(\lambda_{1})\varepsilon_{2}^{\dagger\delta}(\lambda_{2})\left[\varepsilon_{2\alpha}(\lambda_{2})\varepsilon_{3}^{\dagger\mu}(\lambda_{3})\varepsilon_{\boldsymbol{\psi}}^{\dagger\alpha} +\varepsilon_{2}^{\mu}(\lambda_{2})\varepsilon_{3\alpha}^{\dagger}(\lambda_{3})\varepsilon_{\boldsymbol{\psi}}^{\dagger\alpha}-\varepsilon_{2\alpha}(\lambda_{2})\varepsilon_{3}^{\dagger\alpha}(\lambda_{3}) \varepsilon_{\boldsymbol{\psi}}^{\dagger\mu}\right]\nonumber\\
&&\times\left\{ \beta q_{\kappa}\varepsilon_{1\omega}(\lambda_{1})\varepsilon_{3}^{\omega}(\lambda_{3})\varepsilon_{\boldsymbol{\rho}}^{\dagger\kappa}+2\lambda m_{D^{\ast}}p_{1\kappa}\varepsilon_{\boldsymbol{\rho}}^{\dagger\omega}\left[\varepsilon_{1}^{\kappa}(\lambda_{1})\varepsilon_{3\omega}(\lambda_{3})-\varepsilon_{1\omega}(\lambda_{1})\varepsilon_{3}^{\kappa}(\lambda_{3})\right]\right\} ,\label{eq:jrhodast}\\
i\mathcal{M}_{\textbf{(d)}}^{[D]}&=&C_{\textbf{(d)}}^{[D]}\int\frac{d^{4}q}{(2\pi)^{4}}\frac{g^{\alpha\mu}-(p_{1}+q)^{\alpha}(p_{1}+q)^{\mu}/m_{D^{\ast}}^{2}}{(p_{1}+q)^{2}-m_{D^{\ast}}^{2}+i\epsilon} \frac{g_{\alpha\nu}-(p_{2}-q)_{\alpha}(p_{2}-q)_{\nu}/m_{D^{\ast}}^{2}}{(p_{2}-q)^{2}-m_{D^{\ast}}^{2}+i\epsilon}\frac{p_{1\mu}\varepsilon_{\boldsymbol{\chi_{c1}}}^{\dagger\nu}}{q^{2}-m_{D}^{2}+i\epsilon},\label{eq:chic1pid}
\end{eqnarray}
in which we used the notations, such as $\mathcal{M}_{\textbf{(a)}}^{[\pi]}$, where the subscript represents the label of diagram in Fig.~\ref{fig:decays}, and the superscript means the exchanged particle is pion. The $\varepsilon_{\bm\psi}$, $\varepsilon_{\bm\rho}$ and $\varepsilon_{\bm{\chi_{c1}}}$ denote the polarization vectors of $J/\psi$, $\rho$ and $\chi_{c1}$ in order. The corresponding coupling constants from three vertices are packed into the coefficients, such as $C_{\textbf{(a)}}^{[\pi]}$, and their expressions are given in the following,
\begin{align}\label{eq:coeffs}
C_{\textbf{(a)}}^{[\pi]}&=\frac{g_0 g_b^2}{f_\pi^2}m_D m_{D^\ast},&
C_{\textbf{(b)}}^{[D]}&=\frac{2\sqrt{2}g_0 g_2 g_b}{f_\pi}m_D m_{D^\ast}\sqrt{m_{\eta_c}},\nonumber\\
C_{\textbf{(b)}}^{[D^\ast]}&=\frac{2\sqrt{2}g_0 g_2 g_b}{f_\pi}\frac{m_{D^\ast}}{\sqrt{m_{\eta_c}}},&
C_{\textbf{(c)}}^{[D]}&=4 \sqrt{2} g_0 g_2 \lambda g_V \frac{m_D}{\sqrt{m_{\psi}}},&\nonumber\\
C_{\textbf{(c)}}^{[D^\ast]}&=2 \sqrt{2} g_0 g_2 g_V m_{D^\ast}\sqrt{m_{\psi}},&
C_{\textbf{(d)}}^{[D]}&=-\frac{4 g_0 g_1 g_b}{f_\pi} m_D m_{D^\ast} \sqrt{m_{\chi_{c1}}}.&
\end{align}
The expression for $\mathcal{M}_{\textbf{(c)}}^{[D^\ast]}$ is lengthy, so we have written it in the form of Eq.~\eqref{eq:jrhodast}, where its specific form can be obtained by expanding the numerator and then summing over the polarization vectors in the following manner,
\begin{eqnarray}
\sum_{\lambda_1=-1,0,1}\varepsilon_{1\mu}^\dagger(p_1+q,\lambda_1)\varepsilon_{1\nu}(p_1+q,\lambda_1)&=&-g_{\mu\nu}+\frac{(p_1+q)_\mu (p_1+q)_\nu}{m_{D^{\ast}}^2},\\
\sum_{\lambda_2=-1,0,1}\varepsilon_{2\mu}^\dagger(p_2-q,\lambda_2)\varepsilon_{2\nu}(p_2-q,\lambda_2)&=&-g_{\mu\nu}+\frac{(p_2-q)_\mu (p_2-q)_\nu}{m_{\bar{D}^{\ast}}^2},\\
\sum_{\lambda_3=-1,0,1}\varepsilon_{3\mu}^\dagger(q,\lambda_3)\varepsilon_{3\nu}(q,\lambda_3)&=&-g_{\mu\nu}+\frac{q_\mu q_\nu}{m_{D^{\ast}}^2}.
\end{eqnarray}
\end{widetext}
The decay amplitudes for $T_{\psi0}^a\to D\bar{D}$, $\eta_c\pi$, $J/\psi\rho$ and $\chi_{c1}\pi$ are respectively given by
\begin{eqnarray}
\mathcal{M}_{D\bar{D}}&=&\mathcal{M}_{\textbf{(a)}}^{[\pi]},\\
\mathcal{M}_{\eta_c\pi}&=&2\left(\mathcal{M}_{\textbf{(b)}}^{[D]}+\mathcal{M}_{\textbf{(b)}}^{[D^\ast]}\right),\\
\mathcal{M}_{J/\psi\rho}&=&2\left(\mathcal{M}_{\textbf{(c)}}^{[D]}+\mathcal{M}_{\textbf{(c)}}^{[D^\ast]}\right),\\
\mathcal{M}_{\chi_{c1}\pi}&=&2\mathcal{M}_{\textbf{(d)}}^{[D]}.
\end{eqnarray}
With the amplitudes, the partial decay width of $T_{\psi0}^a$ can be expressed as
\begin{eqnarray}
\Gamma_{f_2f_1} &=& \frac{1}{8\pi}\frac{|\bm{p}_1|}{m_{T_{\psi0}^a}^2}\overline{|\mathcal{M}_{f_2f_1}|^2},
\end{eqnarray}
where 
\begin{eqnarray}
|\bm{p}_1|&=&\frac{\sqrt{\mathcal{K}(m_{T_{\psi0}^a}^2,m_{f_1}^2,m_{f_2}^2)}}{2m_{T_{\psi0}^a}},\\
\mathcal{K}(\alpha,\beta,\gamma)&=&\alpha ^2 + \beta ^2 + \gamma ^2-2 \alpha  \beta -2 \alpha  \gamma -2 \beta  \gamma,
\end{eqnarray}
and the overline represents a sum over the polarization(s) of the $\chi_{c1}$ ($J/\psi\rho$) in the final states.

\clabel[r1]{To ensure that the loop integrals in Eqs.~\eqref{eq:ddpi}-\eqref{eq:chic1pid} converge and yield finite results, we employ different form factors to regularize the integrals. Theoretically, the most commonly used form factors can be categorized into three types: i) the Heaviside form factor (step function), ii) the Gaussian form factor, and iii) the multipole form factor. Their expressions are respectively given as
\begin{eqnarray}
    \text{Heaviside form factor: }\mathfrak{F}_1(|\bm p|)&=&\Theta(\Lambda_1-|\bm p|),\label{eq:hff}\\
    \text{Gaussian form factor: }\mathfrak{F}_2(\bm p^2)&=&\exp\left({-\frac{\bm p^2}{\Lambda_2^2}}\right),\label{eq:gff}\\
    \text{Multipole form factor: }\mathfrak{F}_3(p^2)&=&\left(\frac{m_M^2-\Lambda_3^2}{p^2-\Lambda_3^2}\right)^n,\label{eq:mff}
\end{eqnarray}
where $m_M$ denotes the mass of the exchanged particle, and $\Lambda_3=m_M+\alpha_\Lambda\Lambda_{\mathrm{QCD}}$, with $\Lambda_{\mathrm{QCD}}=220$ MeV. $\alpha_\Lambda$ is a dimensionless phenomenological parameter, typically taken to be around $1$~\cite{Wang:2016qmz}. 

We then demonstrate how to regularize the loop integrals in Eqs.~\eqref{eq:ddpi}-\eqref{eq:chic1pid} using these three form factors, beginning with the Heaviside and Gaussian form factors as examples. Since we are dealing with the decays of a bound state, the constituent particles $D^\ast$ and $\bar{D}^\ast$ are off-shell, meaning that there are no right-hand cuts in the loop integrals. Thus, for instance, for a scalar integral with the following form,
\begin{eqnarray} \label{eq:scalarint}
\mathcal{I}=\int\frac{d^{4}q}{(2\pi)^{4}}\frac{1}{(p_{1}+q)^{2}-m_{D^{\ast}}^{2}}\frac{1}{(p_{2}-q)^{2}-m_{\bar{D}^{\ast}}^{2}}\frac{1}{q^{2}-m_{M}^{2}},\nonumber\\
\end{eqnarray}
it can be regularized as
\begin{eqnarray}
\mathcal{I}_i(\Lambda_i)&=&-\frac{3i}{16\pi^{2}}\int_{0}^{1}dx\int_{0}^{1-x}dy\mathscr{I}_i(\Lambda_i).
\end{eqnarray}
The integrands $\mathscr{I}_i(\Lambda_i)$ respectively read
\begin{eqnarray}
\mathscr{I}_1(\Lambda_1)&=&\int_{0}^{\infty}d\ell\frac{\bm{\ell}^{2}\Theta(\Lambda_1-|\bm\ell|)}{\left(\bm{\ell}^{2}+\Delta\right)^{5/2}}
=\frac{\Lambda^{3}_1}{3\Delta\left(\Delta+\Lambda^{2}_1\right)^{3/2}}, \\
\mathscr{I}_2(\Lambda_2)&=&\int_{0}^{\infty}d\ell\frac{\bm{\ell}^{2}e^{-\bm\ell^2/\Lambda_2^2}}{\left(\bm{\ell}^{2}+\Delta\right)^{5/2}}=\frac{\sqrt{\pi}}{4\Delta}U\left(\frac{3}{2},0,\frac{\Delta}{\Lambda_2^2}\right).
\end{eqnarray}
We have used the Feynman parameterization to combine the denominators of the propagators, and employed the residue theorem to integrate out the $\ell_0$ component.
The $\ell=q+xp_{1}-yp_{2}$, and $\Delta=xm_{D^{\ast}}^{2}+y(m_{\bar{D}^{\ast}}^{2}-m_{f_{2}}^{2})+xy(m_{f_{1}}^{2}+m_{f_{2}}^{2}-m_{T_{\psi0}^{a}}^{2})+x^{2}m_{f_{1}}^{2}-xm_{f_{1}}^{2}+y^{2}m_{f_{2}}^{2}-(x+y-1)m_{M}^{2}$ (where $m_M$, $m_{f_1}$ and $m_{f_2}$ denote the masses of the exchanged particles, the final states $f_1$ and $f_2$, respectively). $U(a,b,z)$ represents the Tricomi's (confluent hypergeometric) function.  It can be easily proven that when $\Lambda_{1,2}$ tends to infinity, the result of $\mathcal{I}_i(\Lambda_i)$ is equivalent to that obtained within the dimensional regularization.

Using the Heaviside and Gaussian form factors ensures that all integral terms are convergent. However, when employing multipole form factors, it is necessary to set the power 
$n$ to at least $4$ to guarantee the convergence of all terms in the integral. This choice, however, significantly suppresses the contribution of lower-order terms in $q$. Therefore, we adopt a strategy of incrementally increasing 
$n$ as the power of $q$ in the numerator increases. For example, we multiply the integrand similar to that in Eq.~\eqref{eq:scalarint} by the multipole form factor, and then, by using the Feynman parametrization, we can obtain an integral in the following form,
\begin{eqnarray}
&&\left(m_{M}^{2}-\Lambda_3^2\right)^{n}\frac{\Gamma(3+n)}{\Gamma(n)}\int_{0}^{1}dxdydzdw\delta(x+y+z+w-1)\nonumber\\
&&\qquad\qquad\qquad\qquad\qquad\quad~\times\frac{w^{n-1}\left(\ell^{2}\right)^{n^{\prime}}}{\left(\ell^{2}-\Delta^{\prime}\right)^{3+n}},
\end{eqnarray}
where
\begin{eqnarray}
\begin{cases}
n=1, & n^{\prime}=0,1\\
n=n^{\prime}, & n^{\prime}\in\mathbb{N},n^{\prime}\ge2
\end{cases},
\end{eqnarray}
and
$\Delta^{\prime}=-xy\left(m_{T_{\psi0}^{a}}^{2}-m_{f_{1}}^{2}-m_{f_{2}}^{2}\right)+m_{D^{\ast}}^{2}x+m_{\bar{D}^{\ast}}^{2}y+m_{M}^{2}z+m_{f_{1}}^{2}x^{2}-m_{f_{1}}^{2}x+m_{f_{2}}^{2}y^{2}-m_{f_{2}}^{2}y+\Lambda_{3}^{2}w$.

In Sec.~\ref{sec:nums}, we will demonstrate the effects of these three form factors on the results.
}

\section{Numerical results and discussions}\label{sec:nums}

With the aforementioned preparations, we can now proceed to discuss the dependence of the (partial) decay width of the $T_{\psi0}^a(4010)$ state on the parameters $\Lambda_{1,2}$ and $\alpha_\Lambda$ in the loop integrals, as well as compare the total width with the current experimental data. In Ref.~\cite{Wang:2023hpp}, the mass of the $D^\ast\bar{D}^\ast$ molecular state with quantum numbers $1^-(0^{++})$ was obtained using $\Lambda=0.4$ GeV. In order to maintain consistency with Ref.~\cite{Wang:2023hpp}, we also use $\Lambda=0.4$ GeV in Eq.~\eqref{eq:gft2} to calculate the coupling constant $g_0$. The cutoff parameter $\Lambda$ typically reflects the interaction radius $R$ ($R\sim 1/\Lambda$) of a loosely bound molecular system. Therefore, in principle, the value of $\Lambda_{1,2}$ used in the loop integrals should be consistent with the one used in calculating the mass spectrum. \change{Here, we will vary the value of $\Lambda_{1,2}$ within the range of $0.4$ to $0.7$ GeV to study the dependence of the width on the cutoff, while the $\alpha_\Lambda$ is taken to be in the range of $0.5$ to $1.5$.} 

\begin{figure*}[htbp]
%\begin{centering}
\begin{minipage}[t]{0.33\linewidth}
\centering
\includegraphics[width=\columnwidth]{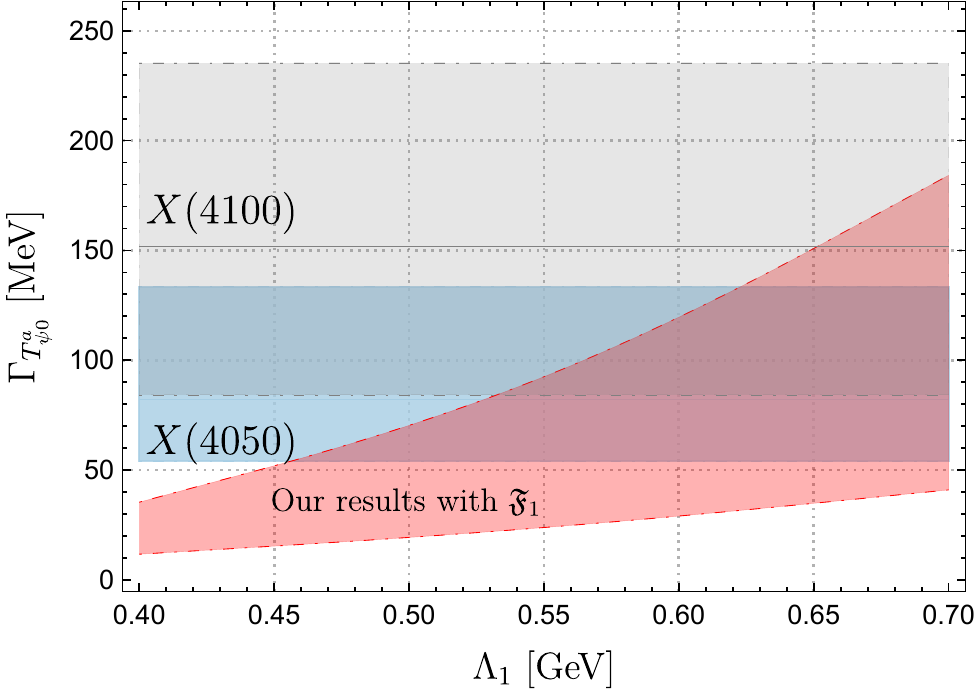}
\end{minipage}%
%\hspace{0.11cm}
\begin{minipage}[t]{0.33\linewidth}
\centering
\includegraphics[width=\columnwidth]{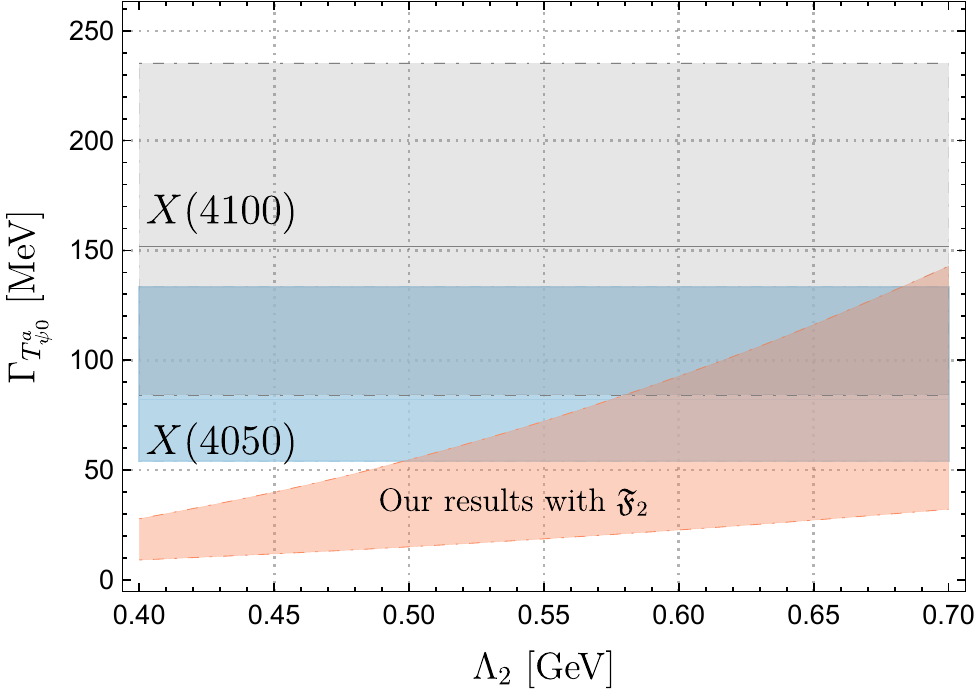}
\end{minipage}
%\hspace{0.1cm}
\begin{minipage}[t]{0.33\linewidth}
\centering
\includegraphics[width=\columnwidth]{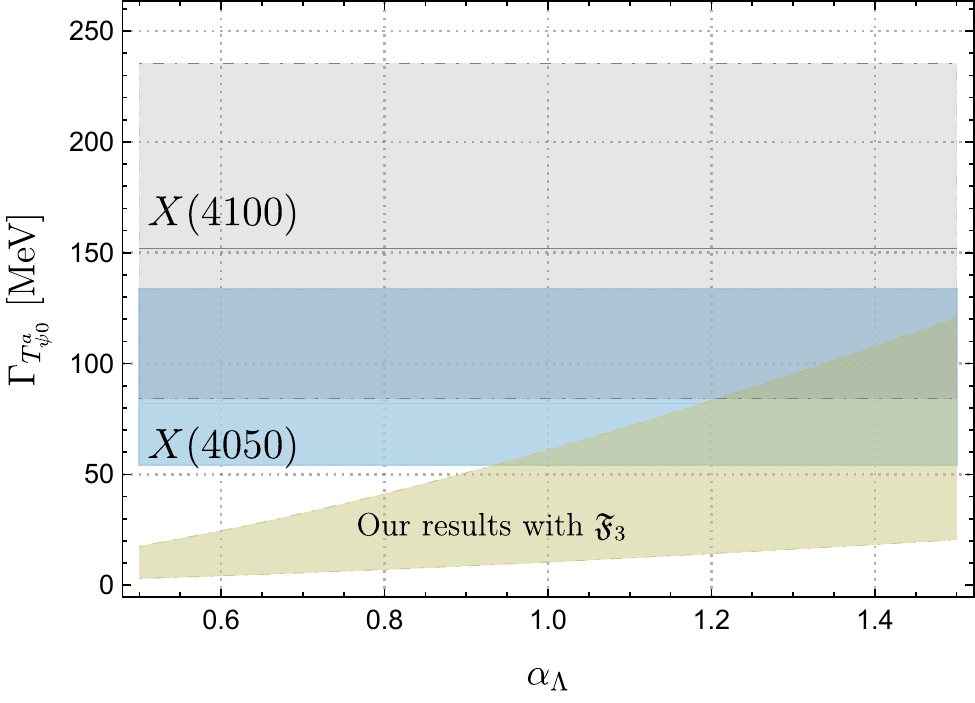}
\end{minipage}
%\scalebox{1.0}{\includegraphics[width=0.95\linewidth]{FitZc.pdf}}
\caption{The dependence of the total decay width of $T_{\psi0}^a(4010)$ on the cutoff parameters. The gray and blue shaded areas represent the widths of $X(4100)$~\cite{LHCb:2018oeg} and $X(4050)$~\cite{Belle:2008qeq}, respectively. The red/orange/yellow bands in the left/middle/right figures represent our results calculated with the Heaviside/Gaussian/Multipole form factors. The range of the band is given by the range of the coupling constant $g_0$ in Eq.~\eqref{eq:g0}. \label{fig:totonlambda}}
%\end{centering}
\end{figure*}

In Fig.~\ref{fig:totonlambda}, we present the dependence of the total width on $\Lambda_{1,2}$ and $\alpha_\Lambda$, and also plot the widths of the $X(4100)$ reported by LHCb~\cite{LHCb:2018oeg} and the $X(4050)$ observed by Belle~\cite{Belle:2008qeq}. These two states have similar masses and widths, and the observed decay channels are precisely the decay modes of $T_{\psi0}^a(4010)$. Therefore, it is possible that they are the same state, serving as candidate for the $1^-(0^{++})$ $D^\ast\bar{D}^\ast$ molecular state. \clabel[r2]{From Fig.~\ref{fig:totonlambda}, it can be observed that the total width increases with increasing $\Lambda_{1,2}$ and $\alpha_\Lambda$. For example, when the mass of $T_{\psi0}^a$ is respectively fixed at $4007.2$ and $4016.7$ MeV, the dependence of the total width on the cutoff parameter can be respectively inferred from the upper and lower boundary curves of the band we calculated.} When $\Lambda_{1,2}$ and $\alpha_\Lambda$ respectively reach around $0.6$ GeV and $1.2$, the theoretically calculated width overlaps with the experimental widths obtained by LHCb and Belle. 

However, it is worth noting that the widths measured by both experiments are obtained using the Breit-Wigner parameterization, which is known to be insufficient in describing the near-threshold states. In such cases, a Flatt\'e-like formula is needed to fit the line shape and extract the pole width. A typical example is the case of $Z_c(3900)$~\cite{BESIII:2013ris} and $T_{cc}(3875)$~\cite{LHCb:2021auc}, where the width obtained using the Flatt\'e-like formula is much smaller than the values obtained using the Breit-Wigner parameterization. Therefore, we suggest that after accumulating more data, the experimental widths be remeasured using the Flatt\'e-like formula in the $\eta_c\pi$ and $\chi_{c1}\pi$ decay channels. \change{Our theoretical estimation for the total width of $T_{\psi0}^a(4010)$ will be provided at $\Lambda_{1,2}=0.4$ GeV and $\alpha_\Lambda=1.0$, and the results are shown in Table~\ref{tab:twidth}.}

\begin{table}[htbp]
\centering
\renewcommand{\arraystretch}{1.5}
\caption{Predictions of the total width of $T_{\psi0}^a(4010)$ within different form factors.\label{tab:twidth}}
\setlength{\tabcolsep}{3.5mm}
{
\begin{tabular}{cccc}
\hline\hline  
Form factor & with $\mathfrak{F}_{1}$ & with $\mathfrak{F}_{2}$ & with $\mathfrak{F}_{3}$\tabularnewline
\hline 
$\Gamma_{T_{\psi0}^{a}}$ ($\mathrm{MeV}$) & $12.0$$-$$35.4$ & $9.0$$-$$27.8$ & $10.5$$-$$61.0$\tabularnewline
\hline\hline  
\end{tabular}
}
\end{table}

From Table~\ref{tab:twidth}, one can see that the width is of similar size with those of its partners, $Z_c(3900)$ and $Z_c(4020)$~\cite{Workman:2022ynf}.

\begin{figure*}[htbp]
%\begin{centering}
\begin{minipage}[t]{\linewidth}
\centering
\includegraphics[width=\columnwidth]{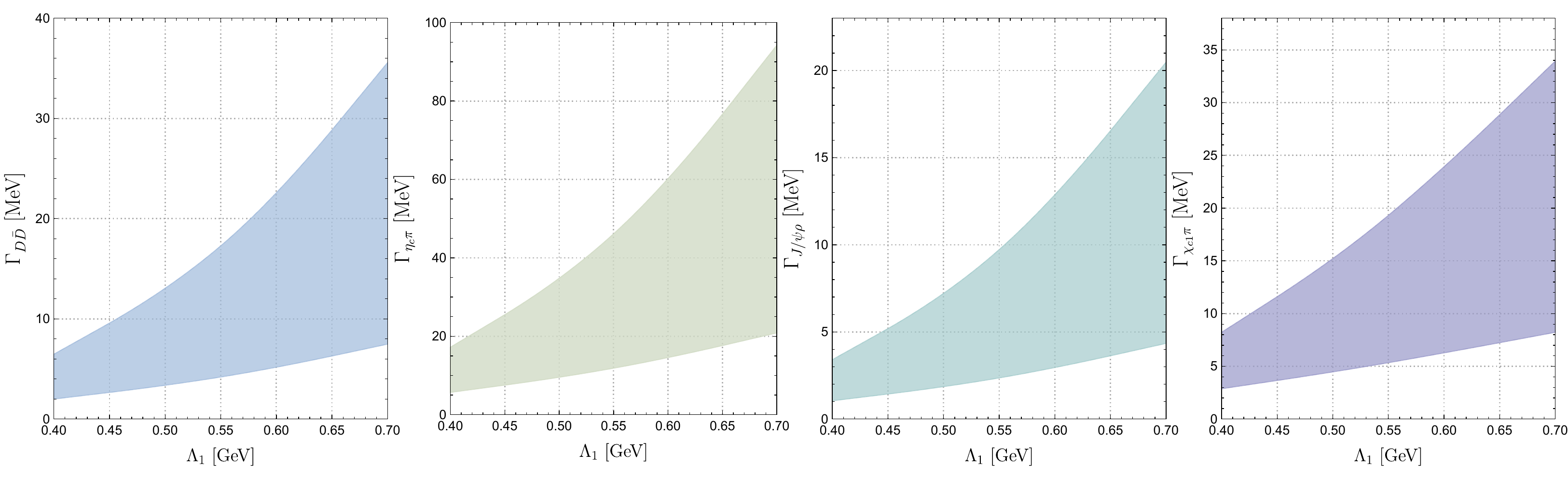}
\end{minipage}%
\\
\begin{minipage}[t]{\linewidth}
\centering
\includegraphics[width=\columnwidth]{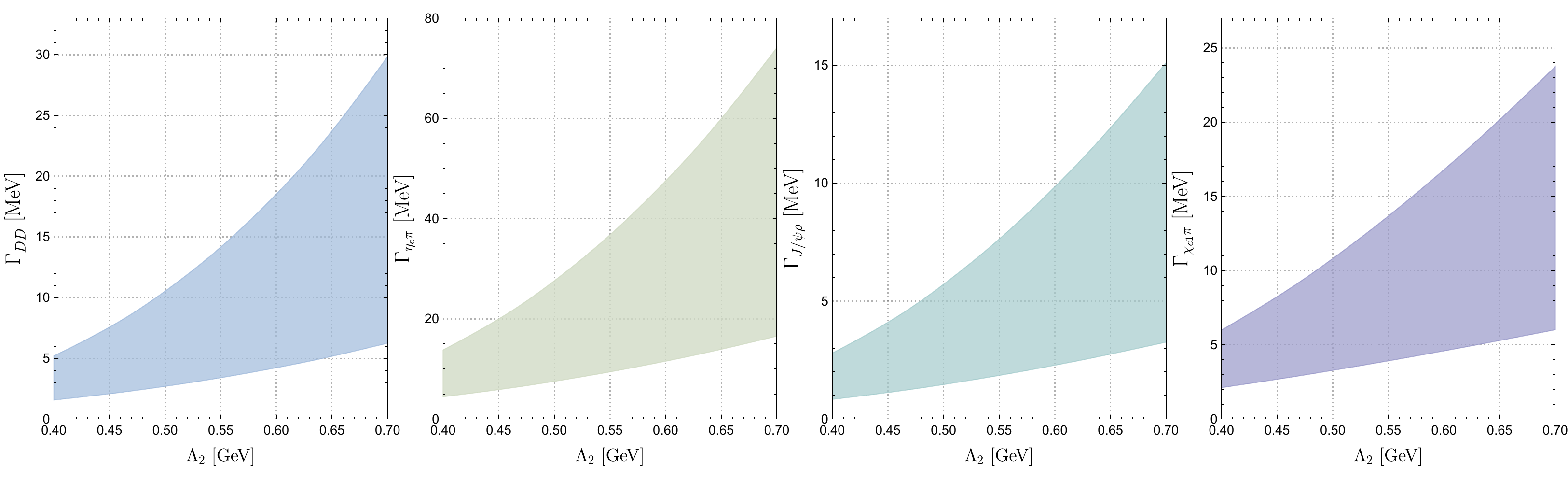}
\end{minipage}
\\
\begin{minipage}[t]{\linewidth}
\centering
\includegraphics[width=\columnwidth]{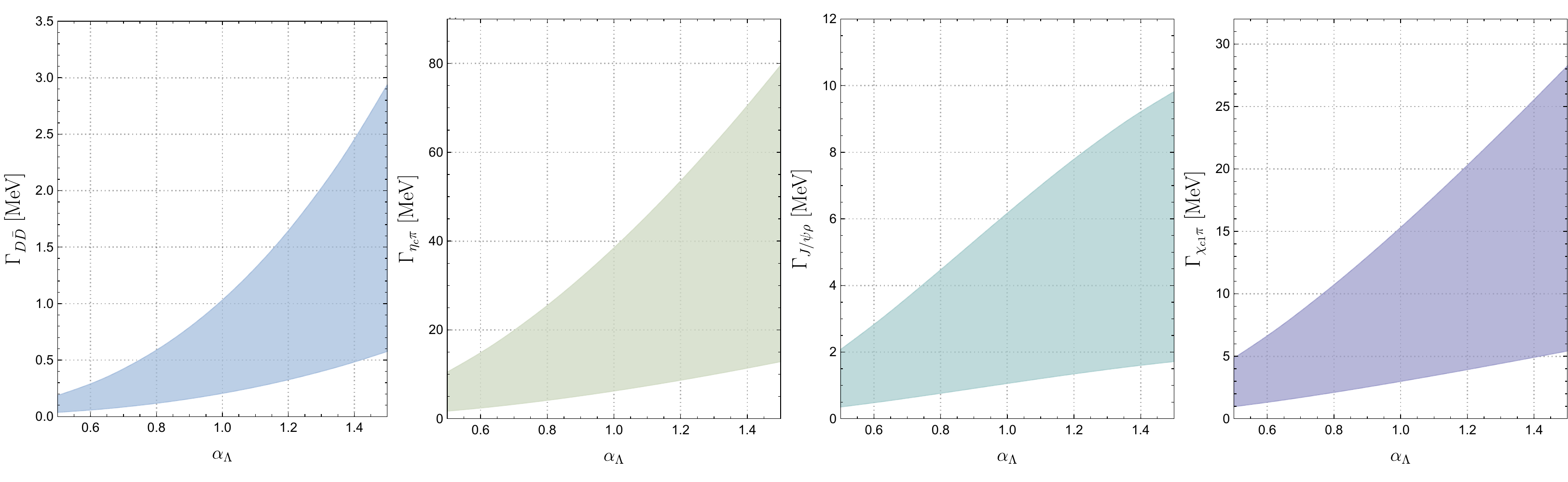}
\end{minipage}
%\scalebox{1.0}{\includegraphics[width=0.95\linewidth]{FitZc.pdf}}
\caption{The dependence of the partial decay widths of $T_{\psi0}^a(4010)$ on the cutoff parameters. The results in the figures of the first/second/third rows represent the calculations with the Heaviside/Gaussian/Multipole form factors.\label{fig:eachwidth}}
%\end{centering}
\end{figure*}

\change{In Fig.~\ref{fig:eachwidth}, we present the dependence of each partial width on $\Lambda_{1,2}$ and $\alpha_\Lambda$, and it can be seen that the partial widths also increase with increasing $\Lambda_{1,2}$ and $\alpha_\Lambda$. Similarly, when $\Lambda_{1,2}$ and $\alpha_\Lambda$ are respectively set to $0.4$ GeV and $1.0$, the range of each partial width is predicted in Table~\ref{tab:eachwidth}.}

\begin{table}[htbp]
\centering
\renewcommand{\arraystretch}{1.5}
\caption{Predictions of the partial decay width of $T_{\psi0}^a(4010)$ within different form factors.\label{tab:eachwidth}}
\setlength{\tabcolsep}{3.7mm}
{
\begin{tabular}{cccc}
\hline\hline 
Form factor & with $\mathfrak{F}_{1}$ & with $\mathfrak{F}_{2}$ & with $\mathfrak{F}_{3}$\tabularnewline
\hline 
$\Gamma_{D\bar{D}}$ ($\mathrm{MeV}$) & $2.0$$-$$6.6$ & $1.6$$-$$5.2$ & $0.2$$-$$1.0$\tabularnewline
%\hline 
$\Gamma_{\eta_{c}\pi}$ ($\mathrm{MeV}$) & $5.8$$-$$17.2$ & $4.5$$-$$13.8$ & $6.2$$-$$38.5$\tabularnewline
%\hline 
$\Gamma_{J/\psi\rho}$ ($\mathrm{MeV}$) & $1.2$$-$$3.4$ & $0.8$$-$$2.8$ & $1.1$$-$$6.2$\tabularnewline
%\hline 
$\Gamma_{\chi_{c1}\pi}$ ($\mathrm{MeV}$) & $3.0$$-$$8.2$ & $2.1$$-$$6.0$ & $3.0$$-$$15.3$\tabularnewline
\hline\hline 
\end{tabular}
}
\end{table}

\begin{figure*}[!ht]
%\begin{centering}
\begin{minipage}[t]{0.33\linewidth}
\centering
\includegraphics[width=\columnwidth]{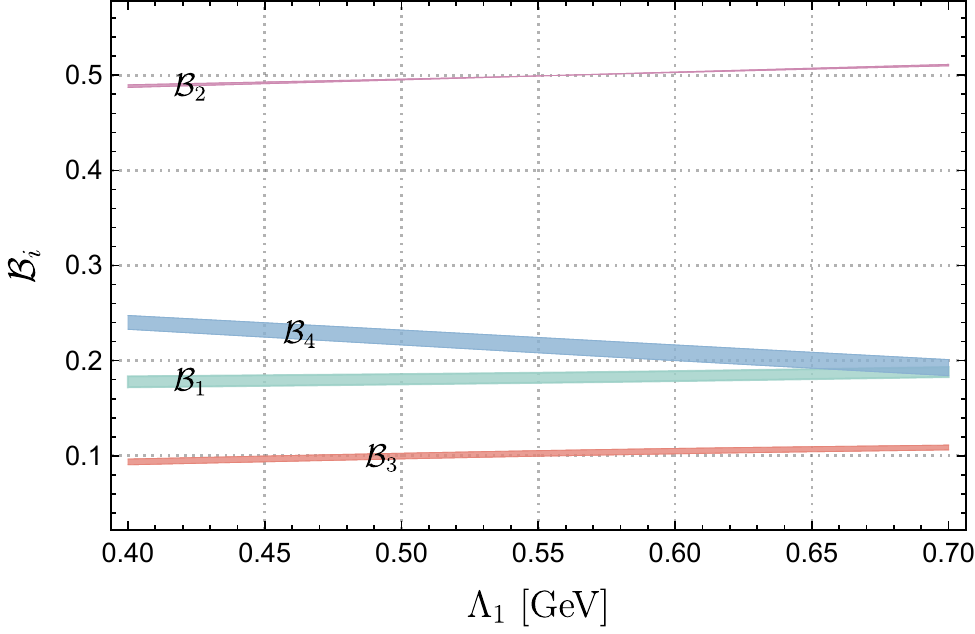}
\end{minipage}%
%\hspace{0.11cm}
\begin{minipage}[t]{0.33\linewidth}
\centering
\includegraphics[width=\columnwidth]{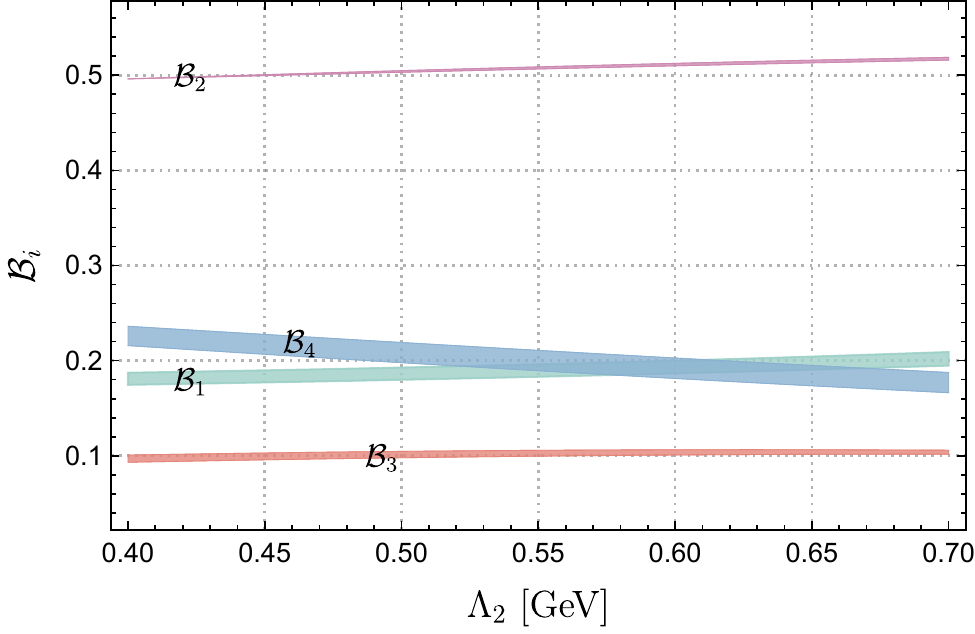}
\end{minipage}
%\hspace{0.1cm}
\begin{minipage}[t]{0.33\linewidth}
\centering
\includegraphics[width=\columnwidth]{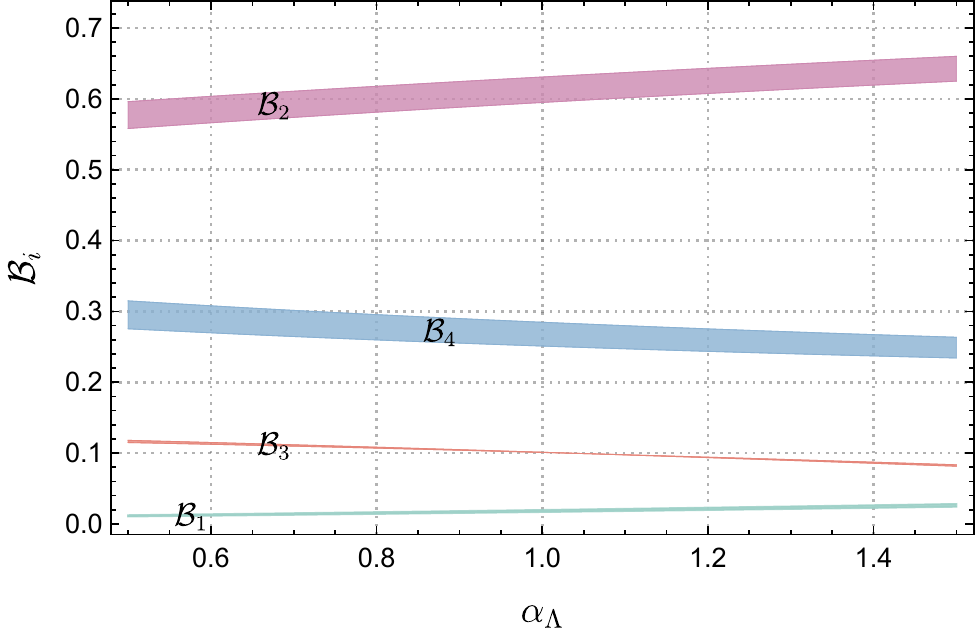}
\end{minipage}
%\scalebox{1.0}{\includegraphics[width=0.95\linewidth]{FitZc.pdf}}
\caption{The dependence of the branching fractions $\mathcal{B}_i$ on the cutoff parameters. The results in the left/middle/right
figures represent the calculations with the Heaviside/Gaussian/Multipole form factors.\label{fig:brafra}}
%\end{centering}
\end{figure*}

\begin{figure*}[!ht]
%\begin{centering}
\begin{minipage}[t]{0.33\linewidth}
\centering
\includegraphics[width=\columnwidth]{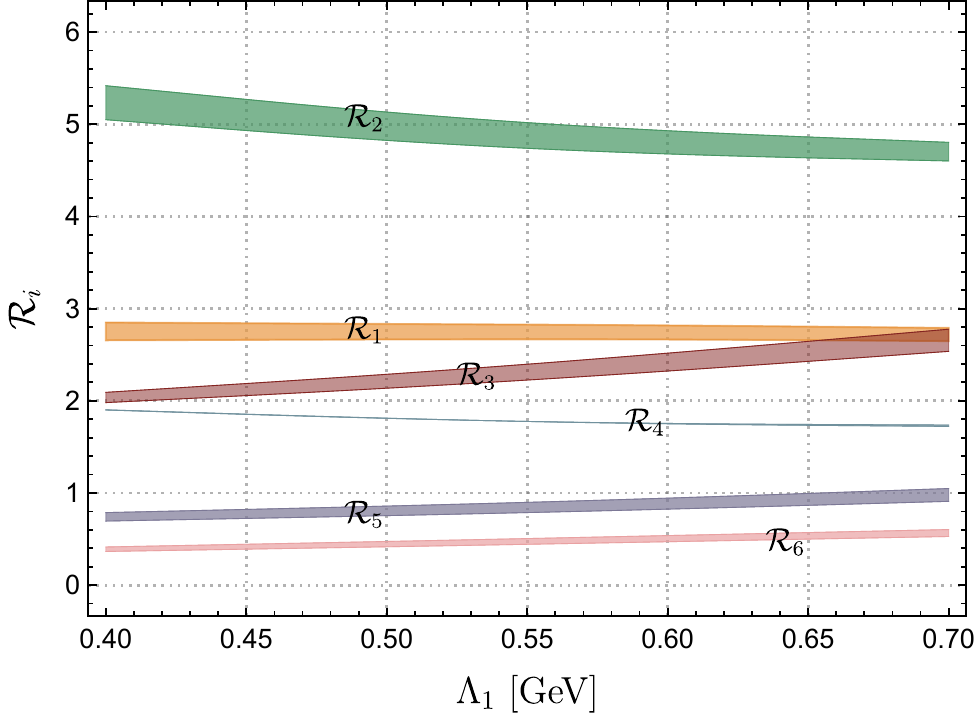}
\end{minipage}%
%\hspace{0.11cm}
\begin{minipage}[t]{0.33\linewidth}
\centering
\includegraphics[width=\columnwidth]{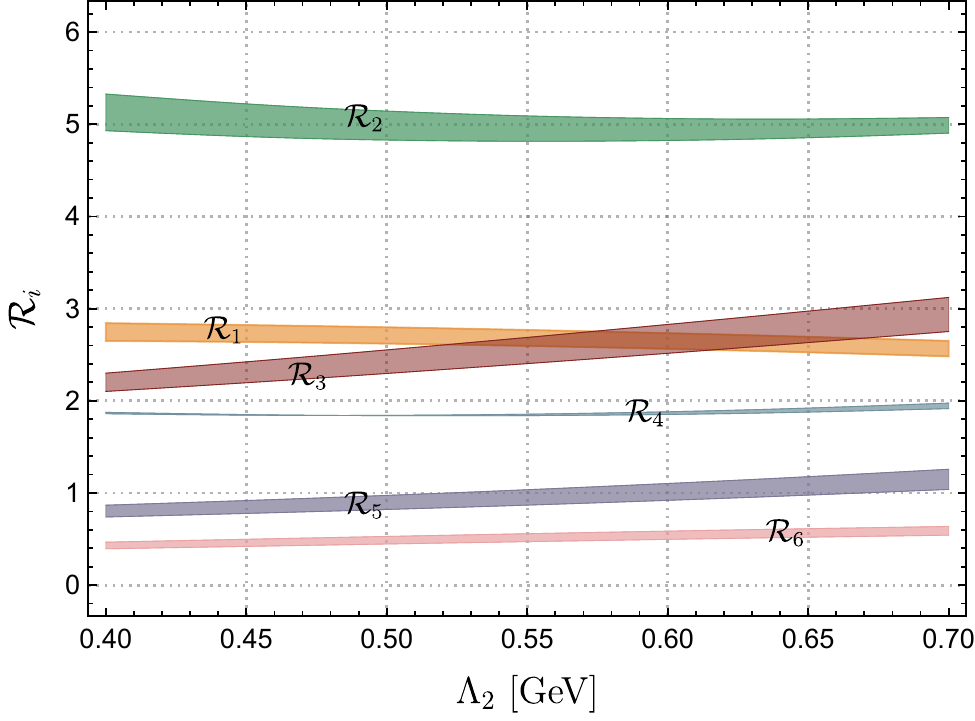}
\end{minipage}
%\hspace{0.1cm}
\begin{minipage}[t]{0.33\linewidth}
\centering
\includegraphics[width=\columnwidth]{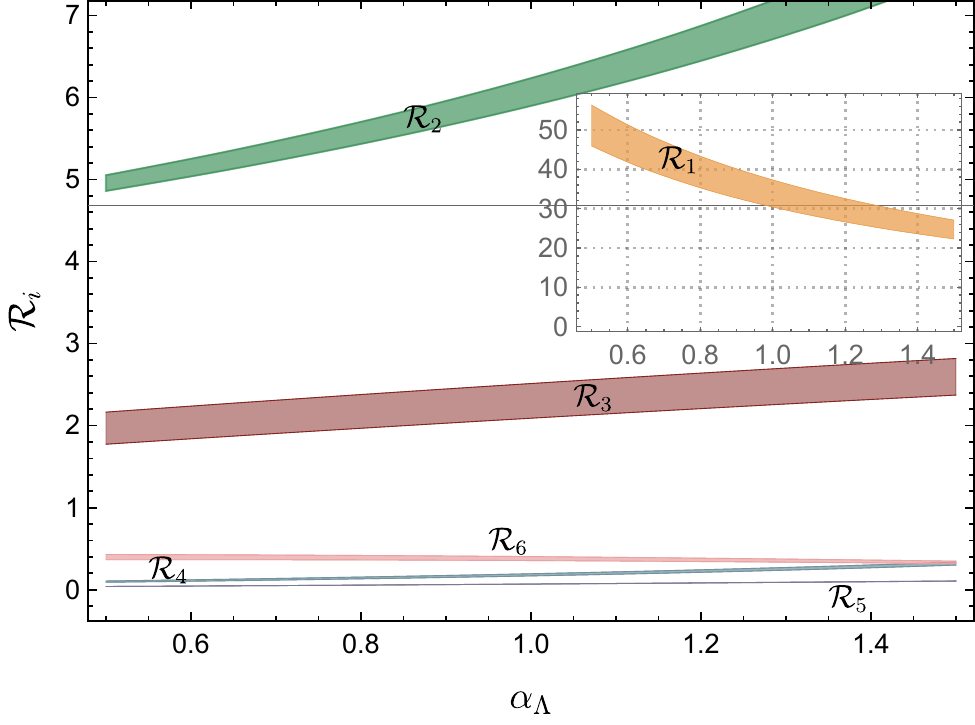}
\end{minipage}
%\scalebox{1.0}{\includegraphics[width=0.95\linewidth]{FitZc.pdf}}
\caption{The dependence of the branching ratios $\mathcal{R}_i$ on the cutoff parameters. The results in the left/middle/right
figures represent the calculations with the Heaviside/Gaussian/Multipole form factors. \label{fig:ratios}}
%\end{centering}
\end{figure*}

\change{The dependence of the branching fractions $\mathcal{B}_i$ and branching ratios $\mathcal{R}_i$ on $\Lambda_{1,2}$ ($\alpha_\Lambda$) are shown in Fig.~\ref{fig:brafra} and Fig.~\ref{fig:ratios}, respectively (The definitions of $\mathcal{B}_i$ and $\mathcal{R}_i$ can be found in the first columns of Table~\ref{tab:brafra} and Table~\ref{tab:ratios}, respectively). From these plots, it can be read that although each partial width is sensitive to the variation of $\Lambda_{1,2}$ ($\alpha_\Lambda$), the dependence of each $\mathcal{B}_i$ and $\mathcal{R}_i$ on $\Lambda_{1,2}$ ($\alpha_\Lambda$) is very weak.} Therefore, in addition to the total width, measuring the $\mathcal{B}_i$ and $\mathcal{R}_i$ experimentally will be very helpful in determining whether $X(4100)$ and $X(4050)$ are the same state and whether they correspond to the $D^\ast\bar{D}^\ast$ molecular state in reality. \change{The predictions for branching fractions and branching ratios are given in Table~\ref{tab:brafra} and Table~\ref{tab:ratios}, respectively, in which we use the average values within the parameter range considering their values exhibit low sensitivity to parameter dependence.}

\begin{table}[htbp]
\centering
\renewcommand{\arraystretch}{1.5}
\caption{Predictions of the branching fractions of $T_{\psi0}^a(4010)$ within different form factors. \label{tab:brafra}}
\setlength{\tabcolsep}{3.9mm}
{
\begin{tabular}{cccc}
\hline\hline 
Form factor & with $\mathfrak{F}_{1}$ & with $\mathfrak{F}_{2}$ & with $\mathfrak{F}_{3}$\tabularnewline
\hline 
$\mathcal{B}_{1}\equiv\Gamma_{D\bar{D}}/\Gamma_{T_{\psi0}^{a}}$  & $0.18$ & $0.19$ & $0.02$\tabularnewline
%\hline 
$\mathcal{B}_{2}\equiv\Gamma_{\eta_{c}\pi}/\Gamma_{T_{\psi0}^{a}}$  & $0.50$ & $0.51$ & $0.61$\tabularnewline
%\hline 
$\mathcal{B}_{3}\equiv\Gamma_{J/\psi\rho}/\Gamma_{T_{\psi0}^{a}}$  & $0.10$ & $0.10$ & $0.10$\tabularnewline
%\hline 
$\mathcal{B}_{4}\equiv\Gamma_{\chi_{c1}\pi}/\Gamma_{T_{\psi0}^{a}}$ & $0.22$ & $0.20$ & $0.27$\tabularnewline
\hline\hline 
\end{tabular}
}
\end{table}

\begin{table}[htbp]
\centering
\renewcommand{\arraystretch}{1.5}
\caption{Predictions of the branching ratios of $T_{\psi0}^a(4010)$ within different form factors.\label{tab:ratios}}
\setlength{\tabcolsep}{3.8mm}
{
\begin{tabular}{cccc}
\hline\hline 
Form factor & with $\mathfrak{F}_{1}$ & with $\mathfrak{F}_{2}$ & with $\mathfrak{F}_{3}$\tabularnewline
\hline 
\hline 
$\mathcal{R}_{1}\equiv\Gamma_{\eta_{c}\pi}/\Gamma_{D\bar{D}}$ & $2.7$ & $2.7$ & $35.4$\tabularnewline
%\hline 
$\mathcal{R}_{2}\equiv\Gamma_{\eta_{c}\pi}/\Gamma_{J/\psi\rho}$ & $4.9$ & $5.0$ & $6.2$\tabularnewline
%\hline 
$\mathcal{R}_{3}\equiv\Gamma_{\eta_{c}\pi}/\Gamma_{\chi_{c1}\pi}$ & $2.3$ & $2.6$ & $2.3$\tabularnewline
%\hline 
$\mathcal{R}_{4}\equiv\Gamma_{D\bar{D}}/\Gamma_{J/\psi\rho}$ & $1.8$ & $1.8$ & $0.19$\tabularnewline
%\hline 
$\mathcal{R}_{5}\equiv\Gamma_{D\bar{D}}/\Gamma_{\chi_{c1}\pi}$ & $0.85$ & $0.96$ & $0.07$\tabularnewline
%\hline 
$\mathcal{R}_{5}\equiv\Gamma_{J/\psi\rho}/\Gamma_{\chi_{c1}\pi}$ & $0.47$ & $0.51$ & $0.37$\tabularnewline
\hline\hline 
\end{tabular}
}
\end{table}

\change{As shown in Tables~\ref{tab:eachwidth},~\ref{tab:brafra} and~\ref{tab:ratios}, the results obtained using the Heaviside and Gaussian form factors are very similar. In contrast, when using the multipole form factor, the width of the $D\bar{D}$ channel is significantly smaller than in the previous two cases, while the widths of the other channels remain relatively close.}
It is evident that the $\eta_c\pi$ channel is the dominant decay mode of the $T_{\psi0}^a$ state, with a decay width approximately twice that of the $\chi_{c1}\pi$ channel. However, the width of $\chi_{c1}\pi$ channel is still larger than the remaining two channels, namely $D\bar{D}$ and $J/\psi\rho$. Therefore, $\eta_c\pi$ and $\chi_{c1}\pi$ are considered the golden channels for detecting the $T_{\psi0}^a$ state. Furthermore, these two channels respectively correspond to the final states in which the $X(4100)$ and $X(4050)$ resonances were observed, which may suggest, to some extent, that $X(4100)$ and $X(4050)$ are molecular state of the $1^-(0^{++})$ $D^\ast\bar{D}^\ast$ system. 

The future experiments can also explore the charged $T_{\psi0}^a$ state by studying the invariant mass spectrum of $D^+\bar{D}^0 +c.c$. For example, with the weak decay processes
$B^0\to D^-D^0K^+$, and $\bar{B}^0\to D^+\bar{D}^0K^-$.
The angular distribution of $D^+\bar{D}^0 +c.c$ will exhibit the characteristic of a flat S-wave distribution. The $D^+\bar{D}^0 +c.c$ channel can effectively exclude contributions from conventional charmonia, such as the $\psi(4040)$. Therefore, the signal observed in the $D^+\bar{D}^0 +c.c$ channel may be cleaner and provide a clearer indication of the presence of the $T_{\psi0}^a$ state. This makes the $D^+\bar{D}^0 +c.c$ channel a promising avenue for studying the $T_{\psi0}^a$ state and distinguishing it from other resonances.

\clabel[r3]{Lastly, it is important to note that our calculations are based on the configuration that $T_{\psi0}^a$ is a bound state of $D^\ast\bar{D}^\ast$. This primarily stems from our previous work~\cite{Wang:2023hpp}, which employed an energy-independent contact potential, where $Z_c(3900)$ is treated as a virtual state. If an energy-dependent contact potential were used, resonance solutions could be obtained~\cite{Albaladejo:2015lob}. However, the current experimental data is insufficient to definitively determine whether $Z_c(3900)$ is a virtual state or a resonance~\cite{Meng:2022ozq}. If $T_{\psi0}^a$ becomes a resonance state of $D^\ast\bar{D}^\ast$, the approaches outlined in this paper could still apply, but the conclusions would change accordingly. For instance, while the coupling constant $g_0$ could still be extracted from the residue of the $T$-matrix at the pole, it would now have an imaginary part due to the resonance mass lying above the $D^\ast\bar{D}^\ast$ threshold. Consequently, the $D^\ast\bar{D}^\ast$ in the loop diagram could be on-shell, and the small imaginary part $i\epsilon$ in the denominator of the propagator cannot be discarded. The resulting scattering amplitude would also contain an imaginary part. In addition to the decay channels listed in this paper, a resonance state could also decay into its constituents, specifically $D^\ast\bar{D}^\ast$. Therefore, the width might be larger than that obtained in the bound state scenario. Moreover, it is very likely that the dominant decay channel is into $D^\ast\bar{D}^\ast$ rather than the hidden-charm decay channel $\eta_c\pi$, as the decay into $D^\ast\bar{D}^\ast$ can occur via tree-level process.}

\section{Summary}\label{sec:sum}

Recently, we investigated the interactions of the $D^{(\ast)}\bar{D}^{(\ast)}$ systems based on a quark-level potential model~\cite{Wang:2023hpp}. We found that if $X(3872)$ and $Z_c(3900)$ are the isoscalar and isovector molecular states of the $D\bar{D}^\ast$ system, respectively, then there must exist a bound state in the $1^-(0^{++})$ $D^\ast\bar{D}^\ast$ system, denoted as $T_{\psi0}^a(4010)$. This state would decay into the $D\bar{D}$, $\eta_c\pi$, $J/\psi\rho$ and $\chi_{c1}\pi$ channels. It is noteworthy that the LHCb and Belle Collaborations have observed the $X(4100)$~\cite{LHCb:2018oeg} and $X(4050)$~\cite{Belle:2008qeq} in the final states of $\eta_c\pi$ and $\chi_{c1}\pi$, respectively. The masses and widths of these two states are of similar size within the experimental uncertainties, and their masses are close to the $D^\ast\bar{D}^\ast$ threshold. Furthermore, the final states of their decays are consistent with the decay channels of $T_{\psi0}^a(4010)$. Therefore, it is significant to investigate whether these two states are the same one and whether they correspond to the molecular state of $D^\ast\bar{D}^\ast$.

In this work, we used the effective Lagrangian approach to investigate the strong decays of the $T_{\psi0}^a(4010)$ state through triangle loop diagrams. We found that its main decay channels are $\eta_c\pi$ and $\chi_{c1}\pi$, which may explain why the LHCb and Belle Collaborations reported the signals of $X(4100)$ and $X(4050)$ in these two channels, respectively. This also suggests that $X(4100)$ and $X(4050)$ might be identified as the $T_{\psi0}^a(4010)$ state. We also investigated the dependence of the total (partial) width(s) and branching fractions (ratios) on the cutoff parameters in the loop integrals. We noticed that the width shows a strong dependence on the cutoff, while the branching fractions (ratios) exhibit a very weak dependence. Therefore, experimental measurements of the branching fractions (ratios) would help to identify the properties of $X(4100)$ and $X(4050)$ and their relationship with the $T_{\psi0}^a(4010)$. \change{Our calculations predict a total width of few tens MeV for the $T_{\psi0}^a(4010)$ within three different form factors}, which is consistent with the experimentally measured widths of its partners $Z_c(3900)$ and $Z_c(4020)$.

We propose that future experiments focus on the study of the resonance parameters of the $T_{\psi0}^a(4010)$ state in the $\eta_c\pi^-$, $\chi_{c1}\pi^-$ and $D^0D^-$ invariant mass spectra of the $B^0\to\eta_c\pi^-K^+$, $\chi_{c1}\pi^- K^+$ and $D^0D^-K^+$ processes (or the charge conjugate channels). This would be crucial for constructing the mass spectrum of the hadronic molecules in the $D^{(\ast)}\bar{D}^{(\ast)}$ systems and, in turn, understanding the properties of the exotic hadrons such as $X(3872)$ and $Z_c(3900)$.

\section*{Acknowledgement}
B. Wang is very grateful to Dr. Lu Meng for helpful discussions. This work is supported by the National Natural Science Foundation of China under Grants No. 12105072. B. Wang is also supported by the Start-up Funds for Young Talents of Hebei University (No. 521100221021).

\bibliography{refs}
\end{document}